\documentstyle[11pt,psfrag,epsfig]{article} 

\begin{document}

\vspace*{2cm}
\begin{center}
{\large{\bf SCALING CONTRIBUTIONS TO THE FREE ENERGY \\ IN THE 1/N
EXPANSION OF
O(N) NONLINEAR \\ SIGMA MODELS IN d-DIMENSIONS}}
\end{center}

\vskip 0.2cm 
\begin{center}
M. Dilaver$^{\star}$\footnote{Hacettepe University, 
Physics Department, 06532, Beytepe, Ankara, Turkey \\ \hspace*{0.5cm} 
{\bf{fax: + (90) 312 2352550; \quad e-mail:
dilaver@thep1.phys.hun.edu.tr}}},
P.
Rossi$^{\star\star}$ and Y. G\"{u}nd\"{u}\c{c}$^{\star}$ \\
\end{center}

$^{\star}${\small Hacettepe University, Physics Department,  06532
Beytepe, Ankara, Turkey. \\ 

 $^{\star\star}${\small Dipartimento di Fisica dell'Universita and INFN ,
I-56100 Pisa, Italy.} \\

{\vskip 0.1cm} 

{\bf Abstract.}

Within the $1/N$ expansion of $O(N)$ nonlinear $\sigma$ models for $d \leq 4$
it is possible to separate consistently the spin-wave and the massive-mode
contributions to the scaling part of the free energy near criticality, and
to evaluate them to $O(1/N)$. For critical dimensions $d = 2+2/n$ the Abe-Hikami
anomaly is recovered, while for $d=2$ the removal of the spin-wave term is justified. 

\vskip 0.5 cm
\newpage

Scaling is a fundamental property of physical systems in the
neighborhood of a second order phase transition.  In the field theory
approach to the study of critical phenomena it is easy to identify the
scaling properties of correlation functions.  However it may be
difficult to study scaling of bulk thermodynamical properties, and in
particular the free energy.  The reason behind this is the phenomenon
of mixing: in a renormalizable field theory the vacuum expectation
values of composite operators require in general not only a
multiplicative renormalization, but also the subtraction of the
contributions coming from lower-dimensionality operators carrying the
same quantum numbers.  These contributions, when present, are
numerically dominant with respect to the scaling part of the
expectation value, for the very simple reason that they are associated
with lower powers of the (vanishing) mass scale, or inverse
correlation length.  Therefore one must in principle compute these
terms with infinite accuracy (summing infinite orders of perturbation
theory) in order for the evaluation of scaling contributions to become
possible.  In practice this "subtraction of perturbative tails" has
been attempted with some success even in purely numerical computations
of topological susceptibilities \cite{A.DiGiacomo:1992}. The reason behind
this
possibility stays in the fact that the physical degrees of fredom
generating these contributions, the "spin waves", are quite different
in nature from the (massive, topological) degrees of freedom
associated with the scaling terms. Therefore, in a numerical
simulation based on heating, the spin waves can be excited quite
independently and earlier than the massive fields. When a plateau is
temporarily reached, the value of the plateau can be nonperturbatively
identified with the spin wave contribution, to be subtracted from the
final (fully thermalized) vacuum expectation value.

There are obvious conceptual and technical limits in the accuracy of this 
determination, and it would be nice to possess some analytical 
nonperturbative scheme in order to deal with this problem.

In the context of two-dimensional $O(N)$ lattice spin models, and of
their regularized continuum counterparts, it was noticed some time ago
that the $1/N$ expansion offered the possibility of consistently
defining the "perturbative tails" of the free and internal energies,
at least up to second nontrivial order \cite{M.Campostrini:1990}.  The
procedure adopted
might however appear to be tailored upon the two-dimensional case,
where criticality occurs in the zero-coupling limit, the models are
perturbatively renormalizable and asymptotically free, and the whole
procedure essentially amounted to a prescription (principal value
integration) for the evaluation of the integral representing the
resummation of an infinite series of perturbative Feynman diagrams.
We therefore decided to consider the problem of evaluating the scaling
part of the free energy in $O(N)$ nonlinear sigma models around
criticality in dimensions $2<d<4$, where criticality occurs at a finite
value of the lattice coupling, and the critical exponents (hence the
scaling properties) are nontrivial.

Finding a subtraction procedure allowing for a nonambiguous identification
of the scaling contributions cannot in this case be considered simply as
the problem of making sense out of the formal sum of a known perturbative
series, since we must first face the problem of correctly identifying
which contributions come from spin wave degrees of freedom.

For definiteness, let's consider the continuum version of nonlinear
sigma models:						

\begin{equation}
{\cal{S}} = \frac{1}{2} N \beta\int
d^{d}x{\partial}_{\mu}\vec{{\mathbf{S}}}.{\partial}_{\mu}\vec{{\mathbf{S}}}
\label{eq1}
\end{equation}

We label the coefficents of the $1/N$ expansion for any physical quantity $Q$ according 
to the notation:

\begin{equation}
Q = \sum_{i}Q_i N^{-i}
\label{eq2}
\end{equation}

The (unsubtracted) free energy of the above model can be computed in the $1/N$
expansion and its formal expression is ;

\begin{equation}
F = \frac{N}{2}\int \frac{d^d p}{(2\pi)^d} \ln \beta (p^2 + m_0^2 ) - 
\frac{N}{2}\beta m_0^2
+ \frac{1}{2}\int \frac{d^d p}{(2\pi)^d}\ln [\Delta^{-1}(p,m_0 )] + O(1/N)
\label{eq3}
\end{equation}

where $m_0$ is defined by the gap equation:

\begin{equation}
\beta = \int \frac{d^d p}{(2\pi)^d}\frac{1}{p^2 + m_0^2}
\label{eq4}
\end{equation}
see ref \cite{Biscari:1990} for explainations and notation.

We introduced the inverse propagator of the effective field:

\begin{eqnarray}
\Delta^{-1}(p,m_0 ) &=& \frac{1}{2}\int \frac{d^d p}{(2\pi)^d}\frac{1}{q^2 + m_0^2}
\frac{1}{(p+q)^2 + m_0^2} \nonumber \\
&=& \frac{\Gamma(2-d/2)}{2(4\pi)^{d/2}} (\frac{p^2}{4} + m_0^2 )^{d/2 -
2}\frac{}{}{ _{2}F_{1}}(2-\frac{d}{2},\frac{1}{2};\frac{3}{2};
\frac{p^2}{p^2 +
m_0^2})
\label{eq5}
\end{eqnarray}
where $_{2}F_{1}(2-\frac{d}{2},\frac{1}{2};\frac{3}{2}; \frac{p^2}{p^2 +
m_0^2})$
is hypergeometric function \cite{I.Gradshteyn}.

It will be essential to our analysis that the inverse propagator admits an asymptotic
expansion for small $m_0$ \cite{V.F.Muller:1986,M.Campostrini:1993}, which
is easily extracted from the
representation:

\begin{equation}
\Delta^{-1}(p,m_0 ) = \frac{\Gamma(2-d/2)}{2(4\pi)^{d/2}} \frac{\Gamma(d/2-1)^2}
{\Gamma(d-2) p} (p^2+4 m_0^2)^{\frac{d-3}{2}}+ \frac{(\beta - \beta_c )}
{p^2+4 m_0^2}\frac{}{}\frac{}{} { _{2}F_{1}}(\frac{d-1}{2},
1;\frac{d}{2};\frac{1}{1+\frac{p^2}{4
m_0^2}})
\label{eq6}
\end{equation}

where $\beta_c$ is the (scheme-dependent) critical value of the coupling and one can show
that, for small $m_0$,

\begin{equation}
(\beta - \beta_c ) = \frac{\Gamma(1-d/2)}{(4\pi)^{d/2}}(m_0^2 )^{d/2 -1}
\label{eq7}
\end{equation}

Such an asymptotic expansion can in turn be reinterpreted as the effect of an operator 
expansion.

In order to identify the spin-wave contribution to the free energy we must evaluate
the small-mass limit of eq.(\ref{eq3}):

\begin{equation}
F^{sw} = \frac{N}{2}\int \frac{d^d p}{(2\pi)^d}\ln(\beta p^2 )+ \frac{1}{2}
\int \frac{d^d p}{(2\pi)^d}\ln\Delta_{0}^{-1}(p,m_0 ) + O(1/N)
\label{eq8}
\end{equation}

where from eq.(\ref{eq6}) we extracted the definition :

\begin{equation}
\Delta_{0}^{-1}(p,m_0) = \frac{\Gamma(2-d/2)\Gamma(d/2-1)^2}{2(4\pi)^{d/2}\Gamma(d-2)}
p^{d-4} + \frac{\Gamma(1-d/2)}{(4\pi)^{d/2}}\frac{m_{0}^{d-2}}{p^2}
\label{eq9}
\end{equation}

The definitions eq.(\ref{eq8}) and eq.(\ref{eq9}) were given with the aim of isolating those
contributions to 
the free energy that are originated from a mixing with the identity operator.
In order to justify our choice we must however prove that the subtracted free
energy possesses the correct scaling properties.

This is achieved by proving finiteness and scheme-independence of the dimensionless ratio;

\begin{equation}
f = \frac{1}{N}(F - F^{(sw)} )m^{-d}
\label{eq10}
\end{equation}

where we have introduced the mass gap $m^2$ (inverse correlation length), whose
large $N$ limit is $m_0^2$. It is easy to obtain the $O(1/N)$ correction in the form:

\begin{equation}
m_1^2 = \int \frac{d^d p}{(2\pi)^d}\frac{\Delta (p,m_0 )}{(p+im_0)^2 + m_0^2}
+ \frac{1}{2} \Delta(0)\int \frac{d^d p}{(2\pi)^d}\Delta (p,m_0 )
\frac{d}{dm_0^2}\Delta^{-1}(p,m_0 )
\label{eq11}
\end{equation}

The ratio $f$ can also be expanded in powers of $1/N$. The calculation of $f_0$
is straight-forward. The result is scheme-independent, and it is most easily derived by
the use of dimensional regularization:

\begin{equation}
f_0 = \frac{1}{2}[\int \frac{d^d p}{(2\pi)^d}\ln (1+\frac{m_0^2}{p^2}) -
\beta m_0^2 ]m_0^{-d} = \frac{2}{d} \Delta^{-1}(0,m_0) m_0^{4-d}
\label{eq12}
\end{equation}

The value of $f_1$ can be obtained from:

\begin{equation}
f_1 = \frac{1}{2}[\int\frac{d^d p}{(2\pi)^d}\ln[\Delta_{0}(p,m_0 )\Delta^{-1}(p,m_0 )]m_0^{-d} - \frac{d}{2}f_0 \frac{m_1^2}{m_0^2}
\label{eq13}
\end{equation}

In order to find a convenient representation for $f_1$, we make the observation
that the total derivative with respect to $m_0^2$ can be  represented in terms
of the partial derivatives according to:

\begin{equation}
\frac{d}{dm_0^2} = \frac{\partial}{\partial m_0^2} + \frac{d\beta}{dm_0^2}
\frac{\partial}{\partial\beta}
= \frac{\partial}{\partial m_0^2} - 2 \Delta^{-1}(0,m_0 )
\frac{\partial}{\partial \beta} 
\label{eq14}
\end{equation}

Moreover the following identity holds under symmetric integration:

\begin{equation}
\frac{1}{(p+im_0 )^2 + m_0^2} \rightarrow \frac{\partial}{\partial \beta}
\Delta^{-1}(p,m_0 ) = \frac{1}{p^2+4 m_0^2}\frac{}{}\frac{}{} {_{2} 
F_{1}}(\frac{d-1}{2},1;\frac{d}{2};\frac{1}
{1+\frac{p^2}{4 m_0^2}})
\label{eq15}
\end{equation}

implying that:

\begin{equation}
m_1^2 = \frac{1}{2} \Delta (0,m_0)\int \frac{d^d p}{(2\pi)^d}\Delta (p,m_0 ) \frac{\partial}{\partial m_0^2} \Delta^{-1}(p,m_0 )
\label{eq16}
\end{equation}

and as a consequence:

\begin{equation}
f_1 = \frac{1}{2}m_0^{-d}\{\int \frac{d^d p}{(2\pi)^d}[\ln\Delta^{-1}(p,m_0 ) - 
\ln \Delta_{0}^{-1}(p,m_0 ) - m_0^2 \frac{\partial}{\partial m_0^2} \ln
\Delta^{-1}(p,m_0 )] \}
\label{eq17}
\end{equation}

We recognize that the expression appearing in the integrand of eq.(\ref{eq17}) must be 
$O(m_0^4 ) $, hence by naive power counting, confirmed by detailed analysis, 
the large-momentum behavior of the integrand turns out to be $O(1/p^4 )$. This
confirms our statement that $f_1$ is finite for all $d<4$, independent of the
regularization scheme adopted.

Let`s notice that $f$ was evaluated for $d=2$ in ref
\cite{M.Campostrini:1990}; and the result was:
$f_0 = \frac{1}{8\pi}$ ,
$f_1 = -2 f_0 $ .

For $d=3$ we obtained: $f_0 = \frac{1}{24\pi}$

\begin{eqnarray}
\frac{f_1}{f_0} &=& \frac{6}{\pi}\int_{0}^{\infty}x^2 dx
[\ln(\frac{2}{\pi}arctan\frac{x}{2}) - 
\ln|1-\frac{4}{\pi x}| - \frac{1}{arctan\frac{x}{2}}(\frac{1}{2}arctan\frac{2}{x} - 
\frac{x}{4+x^2}) ] \\  
&\cong& 1.97863 \nonumber
\label{eq18}
\end{eqnarray}

Since for all $d>1$ $\Delta_{0}^{-1}$ vanishes for some real positive value of $p^2$,
we stress that our integration procedure must always be specified by the request
that no imaginary part should appear in the final result (principal value
prescription ) \cite{M.Campostrini:1993}. Our results for the range $ 1
\le d \le 4 $ are
given by Figure 1.

In order for our analysis to be complete, we must remember that in the critical regime
even the high-frequency modes (spin waves) can bring a scaling contribution to 
the free energy \cite{S.Ma:1974}. While logically distinct from the
quantity we have
just computed
in the context of quantum field theories where a normal-ordering procedure
can be defined in order to remove such a contribution, the scaling free energy
originated from spin waves cannot be isolated in statistical systems and numerical
simulations. Hence it is important to be able to compute the term proportional
to $m^d$ in eq.(\ref{eq8}). This is actually a reasonably simple task, since we can now
exploit the properties of dimensional regularization and evaluate analytically
eq.(\ref{eq8}) for arbitrary $d<1$

From the explicit representation of $\Delta_0^{-1}(p,m_0 )$, eq.(\ref{eq9}), we obtain in
dimensional regularization:

\begin{eqnarray}
F^{(sw)} &=& \frac{1}{2}\int \frac{d^d p}{(2\pi)^d}\ln[p^{d-2} + \frac{4}{2-d}
\frac{\Gamma(d-2)}{\Gamma(d/2-2)^2}m_{0}^{d-2}] \nonumber \\
&=&\alpha_{1}f_0 m_0^d \frac{1}{2(d-1)}[\frac{4}{2-d}\frac{\Gamma(d-2)}
{\Gamma(d/2-1)^2}]^{\frac{2}{d-2}}\Gamma(1-\frac{2}{2-d})\Gamma(\frac{2}{2-d})
\label{eq19}
\end{eqnarray}

where we have introduced the $O(1/N)$ coefficient $\alpha_{1}$ in the expansion
of the critical exponent of the specific heat:

\begin{equation}
\alpha_0 = \frac{d-4}{d-2} \qquad \alpha_1 = \frac{2\Gamma(d)}{\Gamma(2-d/2)\Gamma(d/2)^3}
\label{eq20}
\end{equation}

It is important to notice that the imaginary part of the r.h.s. of eq.(\ref{eq19}) is
exactly needed in order to cancel the imaginary part originated by the naive
integration of eq.(\ref{eq17}), which further justifies the procedure of removing
all imaginary contributions.

Here we face the phenomenon known as Abe-Hikami anomaly \cite{R.Abe:1974
}; whenever
$2/(d-2)$
becomes a positive integer $n$, the expression in eq.(\ref{eq19}) is singular. However
because of eq.(\ref{eq7}), in that same circumstance the expression we obtained is 
proportional to the positive integer power $n+1$ of $\beta - \beta_c$. 
We can therefore subtract a term analytic in $\beta$ from eq.(\ref{eq19}) and obtain
the scaling contribution of spin waves to the free energy in the form:

\begin{equation}
F_{sub}^{(sw)} = \frac{N f_0
m_{0}^{d}}{2(d-1)}[\frac{4}{d-2}\frac{\Gamma(d-2)}
{\Gamma(d/2-1)^2} ]^{\frac{2}{d-2}} + O(1)
\label{eq21}
\end{equation}

The most interesting feature of this result is the dependence on $N$, implying 
that for the special values $d = 2 + 2/n$ even the large $N$ limit of the free energy
is affected by the spin-wave contribution.

When $d \rightarrow 2$, corresponding to $n \rightarrow \infty$, singularities
of the r.h.s. of eq.(20) accumulate on both sides of the real axis in the complex
$d$ plane. However it is easy to check that the limit exists when a small imaginary
part is added to $d$, and its value is consistently $0$, which further justifies 
the procedure of removing all spin-wave contributions in the two-dimensional case.

The support from T\"{U}B\.{I}TAK  is acknowledged.

\newpage
\begin{center}
FIGURE CAPTIONS
\end{center}
Figure 1. $f_1 / f_0$ versus d.

\clearpage
\begin{figure}
\begin{center}
\psfrag{d}{{\huge{\bf{d}}}}
\psfrag{foo}{{\huge{\mbox{\bf{f$_1$ / f$_0$}}}}}
\includegraphics[angle=-90 , width=16cm]{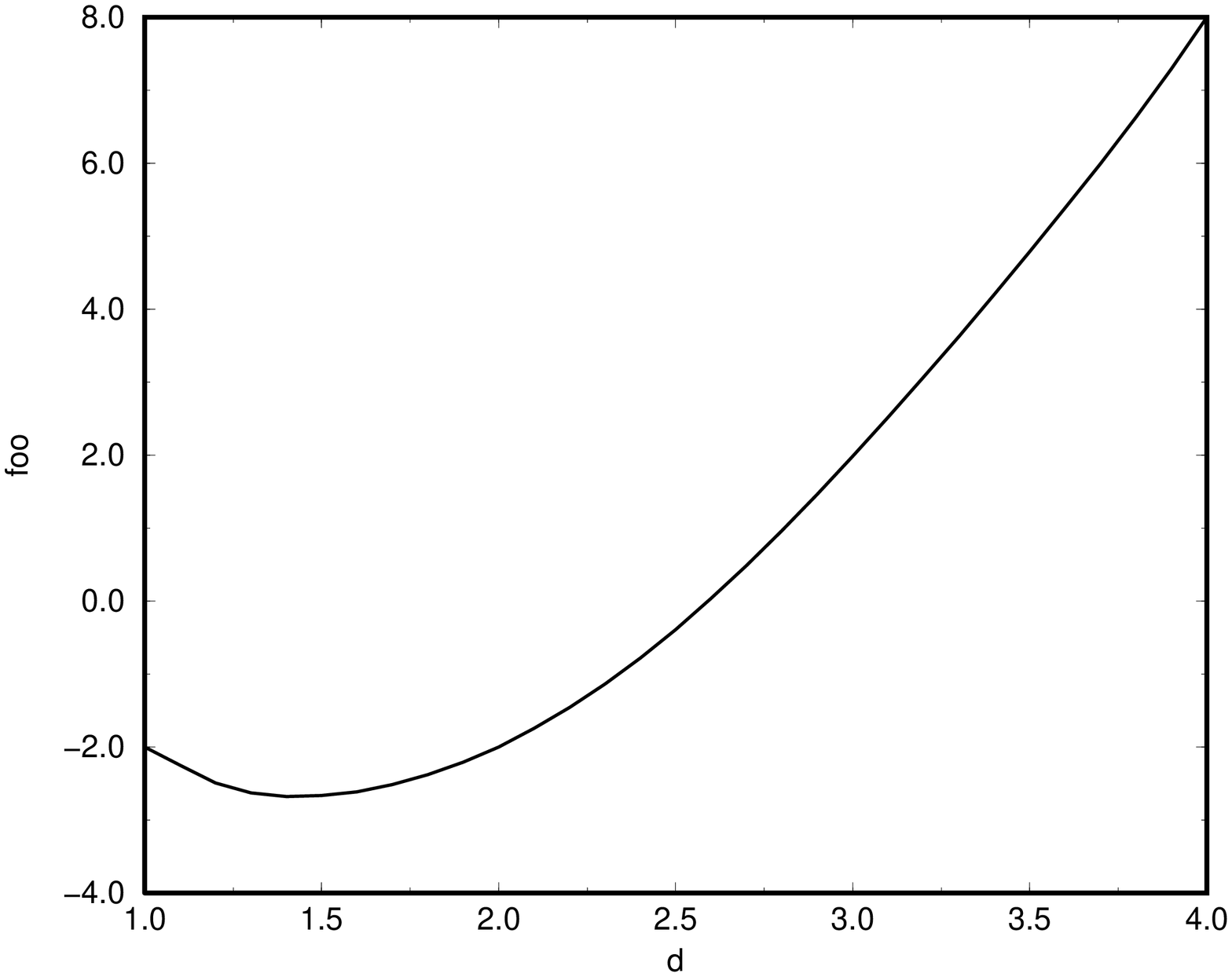}
\end{center}
\caption{}
\end{figure}
\end{document}